\documentclass{aa}

\usepackage{txfonts}
\usepackage{graphicx}
\usepackage{hyperref}
\usepackage{natbib}
\usepackage{epstopdf}
\graphicspath{{./figures/}}

\begin{document}

\title{Statistics of gravitational potential perturbations: \\ 
A novel approach to deriving the X-ray temperature function}
\titlerunning{Statistics of gravitational potential perturbations}

\author{C.~Angrick \and M.~Bartelmann}
\institute{Zentrum f\"ur Astronomie, ITA, Universit\"at Heidelberg, Albert-\"Uberle-Str. 2, 69120 Heidelberg, Germany \\
\email{cangrick@ita.uni-heidelberg.de}}

\date{A\&A accepted; Preprint online version: January 26, 2009}

\abstract
{While the halo mass function is theoretically a very sensitive measure of cosmological models, masses of dark-matter halos are poorly defined, global, and unobservable quantities.}
{We argue that local, observable quantities such as the X-ray temperatures of galaxy clusters can be directly compared to theoretical predictions without invoking masses. We derive the X-ray temperature function directly from the statistics of Gaussian random fluctuations in the gravitational potential.}
{We derive the abundance of potential minima constrained by the requirement that they belong to linearly collapsed structures. We then use the spherical-collapse model to relate linear to non-linear perturbations, and the virial theorem to convert potential depths to temperatures. No reference is made to mass or other global quantities in the derivation.}
{Applying a proper high-pass filter that removes large enough modes from the gravitational potential, we derive an X-ray temperature function that agrees very well with the classical Press-Schechter approach on relevant temperature scales, but avoids the necessity of measuring masses.}
{This first study shows that and how an X-ray temperature function of galaxy clusters can be analytically derived, avoiding the introduction of poorly defined global quantities such as halo masses. This approach will be useful for reducing scatter in observed cluster distributions and thus in cosmological conclusions drawn from them.}


\maketitle

\section{Introduction}

Populating the far end of the halo mass function, galaxy clusters are in principle highly sensitive indicators of the cosmological parameters and non-linear structure growth. Combining Gaussian random density fields with linear structure growth and spherical collapse, the Press-Schechter mass function and its variants turn out to reproduce the halo mass function in fully non-linear cosmological simulations extremely well. If measurable, the abundance of halos in the exponential tail of the mass function and its evolution on cosmic time scales allow precise constraints on both the density-fluctuation amplitude today and during the second half of the cosmic age and on the matter-density parameter. The exponential dependence of the abundance of massive halos on cosmological assumptions promises tight constraints.

A direct comparison between the theoretically predicted mass function of massive halos and the observed distributions of galaxy clusters in various observable quantities, such as the flux and the temperature of their X-ray emission or the velocity dispersion of their member galaxies requires observables to be translated into mass. While this conversion appears straightforward under the idealised assumptions of spherical symmetry, thermal, and hydrostatic equilibrium, the cluster population as a whole shows all signs of being dynamically active. It is doubtful whether precise cosmological conclusions can be drawn based on symmetry assumptions.

Even if clusters satisfied the idealising assumptions typically underlying their cosmological interpretation, their mass is not an observable. In fact, the mass of a dark-matter halo is a poorly defined, derived quantity to which hardly any precise meaning can be given. It is common to operationally define halo masses as enclosed by spheres containing an average fixed overdensity. However, the many different choices of apparently appropriate overdensity values in the literature demonstrate that there is no uniquely defendable choice. If the overdensity is chosen very high, the masses obtained are core masses rather than halo masses, and if it is chosen low, density profiles constrained near the core need to be extrapolated into regions where they are typically poorly measured or not at all.

Halo definitions in numerical simulations illustrate the same problem in a different way. There, halos are typically identified by group finders connecting particles with neighbours closer than a certain linking length. Recipes exist for how the linking length should be chosen, but there is no objective criterion. The dependence on the linking length may be less relevant in practice because halo masses can again be defined as the masses of all particles in spheres containing a fixed overdensity. However, this refers back to the largely arbitrary overdensity threshold and creates the additional problem that several different plausible definitions of halo centres exist that often yield discrepant results.

Three main classes of observation are used to constrain cluster masses: gravitational lensing, X-ray flux and temperature, and galaxy kinematics. None of them measures cluster masses. Gravitational lensing measures the curvature of the projected gravitational potential. X-ray observables are primarily determined by the gas density and temperature, which respond to the depth of the gravitational potential and its gradient, as do galaxy kinematics. Thus, cluster observables constrain the gravitational potential rather than any kind of mass. The conversion of the potential into a mass is hampered by the fact that mass is a non-local quantity, requiring an integration over potential derivatives. We raise the question whether cosmological conclusions can be drawn directly from cluster observables without the detour through problematic definitions of cluster masses.

As one step towards a possible answer, we derive here the X-ray temperature function from a locally defined quantity, namely the gravitational potential. To this purpose, we first derive a function predicting the number density of potential minima having a certain depth. We include the non-linear evolution of the potential by considering the collapse of a spherical and homogeneous overdensity, and locally relate the non-linear potential depth to a temperature using the virial theorem. This direct relation of the temperature to the gravitational potential allows us to avoid introducing a global quantity such as the mass and the ambiguities in its definition. The formalism proposed in this work may thus contribute to reducing the systematic uncertainty in comparisons between theory and observation by avoiding empirical relations between cluster masses and observables.

Unless declared otherwise, we shall use the following cosmological models and parameters: Einstein-de Sitter (EdS): $\Omega_\mathrm{m0}=1.0$, $\Omega_{\Lambda 0}=0.0$, $\Omega_\mathrm{b0}=0.04$, $h=0.7$, $\sigma_8=0.52$; $\Lambda$CDM: $\Omega_\mathrm{m0}=0.3$, $\Omega_{\Lambda 0}=0.7$, $\Omega_\mathrm{b0}=0.04$, $h=0.7$, $\sigma_8=0.93$; OCDM: $\Omega_\mathrm{m0}=0.3$, $\Omega_{\Lambda 0}=0.0$, $\Omega_\mathrm{b0}=0.04$, $h=0.7$, $\sigma_8=0.87$. The different values for $\sigma_8$ reflect the normalisation of the power spectrum to reproduce the local abundance of galaxy clusters \citep{Eke1996}.

\section{Gaussian random fields}

Simple models of inflation predict that the density contrast, $\delta(\vec{r})=\rho(\vec{r})/\rho_{\rm b}-1$, where $\rho(\vec{r})$ is the actual density at position $\vec{r}$ and $\rho_{\rm b}$ is the mean cosmic background density, should be a Gaussian random field right after inflation. Since the density contrast and the gravitational potential are linearly related, the latter is then also a Gaussian random field.

In this section, we shall follow the line of argument presented by \cite{Bardeen1986}. For better comparison to their paper, we adopt $F(\vec{r})$ for the random field and $\vec{\eta}(\vec{r})=\nabla F(\vec{r})$ and $\zeta_{ij}(\vec{r})=\partial_i \partial_j F(\vec{r})$ for its first and second derivatives, respectively.

\subsection{Definition}

An \emph{$n$-dimensional random field} $F(\vec{r})$ assigns a set of random numbers to each point in $n$-dimensional space. A joint probability function can be declared for $m$ arbitrary points $\vec{r}_j$ as the probability that the field $F$, considered at the points $\vec{r}_j$, has values between $F(\vec{r}_j)$ and $F(\vec{r}_j)+{\rm d}F(\vec{r}_j)$ with $j=1,\ldots,m$.

A \emph{Gaussian random field} is a field whose joint probability functions are multivariate Gaussians. Let $y_i$ with $i=1,\ldots,p$ be a set of Gaussian random variables with means $\langle y_i\rangle$ and $\Delta y_i:=y_i-\langle y_i\rangle$. The \emph{covariance matrix} $\tens{M}$ has the elements $M_{ij}:=\langle\Delta y_i \Delta y_j\rangle$, and the joint probability function of the Gaussian random variables is
\begin{equation}
  P(y_1,\ldots,y_p)\,{\rm d}y_1\cdots{\rm d}y_p=
  \frac{1}{\sqrt{\left(2\pi\right)^p\det\left(\tens{M}\right)}}\,
  {\rm e}^{-Q}\,{\rm d}y_1\cdots{\rm d}y_p
\end{equation}
with the quadratic form
\begin{equation}
\label{eq:quadForm}
  Q:=\frac{1}{2}\sum_{i,j=1}^{p}\Delta y_i\left(\tens{M}^{-1}\right)_{ij}\Delta y_j\;.
\end{equation}
A homogeneous Gaussian random field with zero mean is fully characterised by its two-point correlation function $\xi(\vec{r}_1,\vec{r}_2)=\xi(|\vec{r}_1-\vec{r}_2|):=\langle F(\vec{r}_1)F(\vec{r}_2)\rangle$ or equivalently its Fourier transform, the power spectrum $P(k)$.

\subsection{The minimum constraint}

An expression for the number density of field minima can be derived as follows. The joint probability function for a Gaussian random field in three-dimensional space with zero mean including first and second derivatives reads
\begin{equation}
\label{eq:jointProb}
  p(F,\vec{\eta},\tens{\zeta})\,\mathrm{d}F\,\mathrm{d}^3\eta\,\mathrm{d}^6 \zeta=
  \frac{1}{\sqrt{\left(2\pi\right)^{10}\det\left(\tens{M}\right)}}\,
  \mathrm{e}^{-Q}\,\mathrm{d}F\,\mathrm{d}^3\eta\,\mathrm{d}^6\zeta\;,
\end{equation}
with the quadratic form $Q$ given in Eq.~(\ref{eq:quadForm}) and  $\vec{y}=\left(F, \eta_1, \eta_2, \eta_3, \zeta_{11}, \zeta_{22}, \zeta_{33}, \zeta_{12}, \zeta_{13}, \zeta_{23}\right)$.

The matrix $\tens{M}$ contains all auto- and cross-correlations between these quantities, which read
\begin{equation}
\label{eq:correlations}
  \begin{array}{rclcrcl}
    \langle FF\rangle&=&\sigma_0^2\;,&&
    \langle\eta_i\eta_j\rangle&=&\displaystyle{\frac{\sigma_1^2}{3}}\delta_{ij}\;,\\[1.5ex]
    \langle F\zeta_{ij}\rangle&=&-\displaystyle{\frac{\sigma_1^2}{3}}\delta_{ij}\;,&&
    \langle\zeta_{ij}\zeta_{kl}\rangle&=&\displaystyle{\frac{\sigma_2^2}{15}}
    \left(\delta_{ij}\delta_{kl}+\delta_{ik}\delta_{jl}+\delta_{il}\delta_{jk}\right)\;,\\[1.5ex]
    \langle F\eta_{ij}\rangle&=&0\;,&&
    \langle\eta_i\zeta_{jk}\rangle&=&0\;,
  \end{array}
\end{equation}
where the $\sigma_j$, $0\le j\le2$, are the \emph{spectral moments} of the power spectrum $P(k)$,
\begin{equation}
\label{eq:specMoments}
  \sigma_j^2:=\int\frac{k^2 \mathrm{d}k}{2\pi^2}P(k)k^{2j}\hat{W}_R^2(k)\;.
\end{equation}
The Fourier transform $\hat{W}_R(k)$ of the top-hat window function with the filtering scale $R$ is
\begin{equation}
\label{eq:topHat}
  \hat{W}_R(k)=\frac{3\left(\sin u-u\cos u\right)}{u^3}\quad\mbox{with}\quad u=kR\;.
\end{equation}

Let $\vec{r}_0$ be a minimum of the field $F$, hence $\vec{\eta}(\vec{r}_0)=\vec{0}$, and the eigenvalues $(\tilde{\zeta}_1$, $\tilde{\zeta}_2$, $\tilde{\zeta}_3$) of the tensor $(\zeta_{ij})$ of second derivatives must be \emph{positive}. Within an infinitesimal volume $\mathrm{d}^3 r$ around $\vec{r}=\vec{0}$, we can approximate $\eta_i\approx\zeta_{ij} r_j$ and thus replace $\mathrm{d}^3\eta=|\det(\tens{\zeta})| \mathrm{d}^3 r$ in Eq.~(\ref{eq:jointProb}). We also transform the volume element in the space of second derivatives, $\mathrm{d}^6\zeta$, into the space of eigenvalues,
\begin{equation}
  \mathrm{d}^6\zeta=\frac{\pi^2}{3}\left|
    \left(\tilde{\zeta}_1-\tilde{\zeta}_2\right)
    \left(\tilde{\zeta}_2-\tilde{\zeta}_3\right)
    \left(\tilde{\zeta}_1-\tilde{\zeta}_3\right)
  \right|\mathrm{d}\tilde{\zeta}_1\mathrm{d}\tilde{\zeta}_2\mathrm{d}\tilde{\zeta}_3
\end{equation}
\citep{Bardeen1986}. Using $\det(\zeta)=\tilde{\zeta}_1\tilde{\zeta}_2\tilde{\zeta}_3$, we arrive at the final equation for the number density of minima,
\begin{eqnarray}
\label{eq:minima}
  n(F)&=&\frac{\pi^2}{3}
  \int\limits_0^\infty\mathrm{d}\tilde{\zeta}_1
  \int\limits_0^\infty\mathrm{d}\tilde{\zeta}_2
  \int\limits_0^\infty\mathrm{d}\tilde{\zeta}_3\left|
    \tilde{\zeta}_1\tilde{\zeta}_2\tilde{\zeta}_3
  \right|\\&\times&\left|
    \left(\tilde{\zeta}_1-\tilde{\zeta}_2\right)
    \left(\tilde{\zeta}_2-\tilde{\zeta}_3\right)
    \left(\tilde{\zeta}_1-\tilde{\zeta}_3\right)
  \right|\,p\left(F,\vec{\eta}=\vec{0},\tilde{\zeta}_1,\tilde{\zeta}_2,\tilde{\zeta}_3\right)\;.
  \nonumber
\end{eqnarray}

\subsection{Number density of potential minima}

We now apply this formalism to the Gaussian random field of gravitational-potential fluctuations $\Phi$. We continue using $\vec{\eta}$ for the first derivative of the field and $\tens{\zeta}$ for its tensor of second derivatives, but introduce new variables. Instead of the eigenvalues $\tilde{\zeta}_i$ with $i=1,2,3$, we switch to the linear combinations
\begin{equation}
\label{eq:newVar}
  \Delta\Phi:=\tilde{\zeta}_1+\tilde{\zeta}_2+\tilde{\zeta}_3\;,\quad
  \tilde{x}:=\frac{\tilde{\zeta}_1-\tilde{\zeta}_3}{2}\;,\quad
  \tilde{y}:=\frac{\tilde{\zeta}_1-2\tilde{\zeta}_2+\tilde{\zeta}_3}{2}\;.
\end{equation}
These choices simplify the correlation matrix $\tens{M}$, and we can later easily identify the Laplacian of the field. In these new variables, the non-vanishing correlations from Eq.~(\ref{eq:correlations}) are
\begin{equation}
  \begin{array}{rclcrcl}
    \langle\Phi \Delta\Phi\rangle&=&-\sigma_1^2\;,&\quad&
    \langle\tilde{x}\tilde{x}\rangle&=&\displaystyle{\frac{\sigma_2^2}{15}}\;,\\[1.5ex]
    \langle\Delta\Phi\Delta\Phi\rangle&=&\sigma_2^2\;,&\quad&
    \langle\tilde{y}\tilde{y}\rangle&=&\displaystyle{\frac{\sigma_2^2}{5}}\;.
  \end{array}
\end{equation}
The determinant of the covariance matrix then becomes
\begin{equation}
  \det(\tens{M})=\frac{\sigma_1^6\sigma_2^{10}\gamma}{6834375}
  \quad\mbox{with}\quad\gamma:=\sigma_0^2\sigma_2^2-\sigma_1^4\;,
\end{equation}
and the quadratic form, Eq.~(\ref{eq:quadForm}), turns into
\begin{eqnarray}
  Q&=&\frac{3\vec{\eta}\cdot\vec{\eta}}{2\sigma_1^2}+
  \frac{15\tilde{x}^2}{2\sigma_2^2}+\frac{5\tilde{y}^2}{2\sigma_2^2}+
  \frac{15(\zeta_{12}^2+\zeta_{13}^2+\zeta_{23}^2)}{2\sigma_2^2}\nonumber\\&+&
  \frac{\sigma_0^2(\Delta\Phi)^2}{2\gamma}+
  \frac{2\sigma_1^2\Phi \Delta\Phi}{2\gamma}+\frac{\sigma_2^2\Phi^2}{2\gamma}\;.
\end{eqnarray}

In order to find the number density of potential minima, we have to invert the relations given in Eq.~(\ref{eq:newVar}), considering that only the diagonal elements of the tensor $\tens{\zeta}$ are non-zero after transforming to principal axes. After replacing $(\tilde{\zeta}_1, \tilde{\zeta}_2, \tilde{\zeta}_3)$ by $(\Delta\Phi, \tilde{x}, \tilde{y})$ and changing the integration boundaries accordingly, we integrate only over $\tilde{x}$ and $\tilde{y}$ because the Laplacian of the potential will become crucial in the following discussion, when another constraint on $\Delta\Phi$ will be introduced. We can now rewrite Eq.~(\ref{eq:minima}) as
\begin{equation}
  \tilde{n}(\Phi,\Delta\Phi)\mathrm{d}\Phi\mathrm{d}(\Delta\Phi)=
  C(N_1+N_2)\mathrm{d}\Phi\mathrm{d}(\Delta\Phi)\;,
\end{equation}
with the integrals
\begin{eqnarray}
\label{eq:N1}
  N_1&=&\int\limits_{-\Delta\Phi/2}^0\mathrm{d}\tilde{x}
  \int\limits_{-3\tilde{x}-\Delta\Phi}^{\Delta\Phi/2} \mathrm{d}\tilde{y}\left|
    \left(\tilde{x}^3-\tilde{x}\tilde{y}^2\right)\left(\Delta\Phi-2\tilde{y}\right)\left(\Delta\Phi-3\tilde{x}+\tilde{y}\right)
  \right|\nonumber\\&\times&\left|
    \left(\Delta\Phi+3\tilde{x}+\tilde{y}\right)
  \right|\mathrm{e}^{-\tilde{Q}}\;,\\
\label{eq:N2}
  N_2&=&\int\limits_0^{\Delta\Phi/2} \mathrm{d}\tilde{x}
  \int\limits_{3\tilde{x}-\Delta\Phi}^{\Delta\Phi/2} \mathrm{d}\tilde{y}\left|
    \left(\tilde{x}^3-\tilde{x}\tilde{y}^2\right)\left(\Delta\Phi-2\tilde{y}\right)\left(\Delta\Phi-3\tilde{x}+\tilde{y}\right)
  \right|\nonumber\\&\times&\left|
    \left(\Delta\Phi+3\tilde{x}+\tilde{y}\right)
  \right|\mathrm{e}^{-\tilde{Q}}\;,
\end{eqnarray}
the normalisation constant
\begin{equation}
  C=\frac{25\sqrt{5}}{16\pi^3\sigma_1^3\sigma_2^5\sqrt{3\gamma}}
\end{equation}
and the quadratic form
\begin{equation}
  \tilde{Q}=\frac{1}{2}\left[
    \frac{15\tilde{x}^2}{\sigma_2^2}+\frac{5\tilde{y}^2}{\sigma_2^2}+
    \frac{\sigma_0^2(\Delta\Phi)^2}{\gamma}+
    \frac{2\sigma_1^2\Phi \Delta\Phi}{\gamma}+\frac{\sigma_2^2\Phi^2}{\gamma}
  \right]\;.
\end{equation}

Equations~(\ref{eq:N1},\ref{eq:N2}) can be integrated analytically, giving identical results. The final expression for $\tilde{n}(\Phi,\Delta\Phi)$ is
\begin{eqnarray}
\label{eq:numDensPhiDelPhi}
  \tilde{n}(\Phi,\Delta\Phi)&=&\frac{1}{240\pi^3\sigma_1^3\sqrt{15\gamma}}
  \exp\left[-\frac{\left(2\sigma_1^2\Delta\Phi+\sigma_2^2\Phi\right)\Phi}{2\gamma}\right]\nonumber \\ 
&\times&(F_1+F_2)\;,
\end{eqnarray}
where $F_1$ and $F_2$ are functions depending only on the field's Laplacian, but not on the field itself,
\begin{eqnarray}
  F_1&=&2\sigma_2\left(5\Delta\Phi^2-16\sigma_2^2\right)
  \exp\left[
    -\frac{\left(6\sigma_0^2\sigma_2^2-5\sigma_1^4\right)\Delta\Phi^2}{2\sigma_2^2\gamma}
  \right]\nonumber\\&+&
  \sigma_2\left(155\Delta\Phi^2+32\sigma_2^2\right)
  \exp\left[
    -\frac{\left(9\sigma_0^2\sigma_2^2-5\sigma_1^4\right)\Delta\Phi^2}{8\sigma_2^2\gamma}
  \right]\;,\\
\label{eq:numDensPhiDelPhiLast}
  F_2&=&5\sqrt{10\pi}\Delta\Phi\left(\Delta\Phi^2-3\sigma_2^2\right)\exp\left(-\frac{\sigma_0^2\Delta\Phi^2}{2\gamma}\right) \nonumber \\
&\times&\left[\mathrm{erf}\left(\frac{\sqrt{5}\Delta\Phi}{2\sqrt{2}\sigma_2}\right)+
  \mathrm{erf}\left(\frac{\sqrt{5}\Delta\Phi}{\sqrt{2}\sigma_2}\right)\right]\;.
\end{eqnarray}
We point out that Eqs.~(\ref{eq:numDensPhiDelPhi}--\ref{eq:numDensPhiDelPhiLast}) are valid in this form only for $\Delta\Phi>0$ and $\Phi<0$ because the underlying integrations over $\tilde{x}$ and $\tilde{y}$ were carried out under these restrictions. Both assumptions are appropriate; the first because of Poisson's equation, and the second because we are only interested in gravitationally bound objects whose potentials must be negative.

For the further evaluation of Eqs.~(\ref{eq:numDensPhiDelPhi}--\ref{eq:numDensPhiDelPhiLast}), we need the first three spectral moments of the potential power spectrum, defined in Eq.~(\ref{eq:specMoments}).

\section{Linear and non-linear evolution of gravitational fluctuations}

The potential power spectrum $P_\Phi(k)$ is related to the density power spectrum $P_\delta(k)$ through Poisson's equation. The power spectrum, however, only describes the linear evolution of fluctuations for which $\delta\la 1$. Thus, we also need an \textit{ansatz} for their non-linear evolution having higher amplitude. We shall use the \emph{spherical-collapse model} (SCM) to model non-linear effects. Along the way, we shall introduce a proper definition of a filtering scale $R$.

\subsection{Linear power spectrum}

The gravitational potential is related to the density contrast in comoving coordinates by Poisson's equation
\begin{equation}
\label{eq:relateSpectra}
  \Delta\Phi=4\pi G \rho_\mathrm{b} a^2 \delta\;,\quad
  -k^2\hat{\Phi}=4\pi G \rho_\mathrm{b} a^2 \hat{\delta}\;,
\end{equation}
in real and Fourier space, respectively. By the definition of the power spectrum,
\begin{equation}
  \langle\hat{\delta}(\vec{k})\hat{\delta}^\ast(\vec{k}^\prime)\rangle=:
  (2\pi)^3P(k)\delta_\mathrm{D}(\vec{k}-\vec{k}^\prime)\;,
\end{equation}
where $\delta_\mathrm{D}$ denotes \emph{Dirac's delta distribution}, and using $\rho_\mathrm{b}=(3H_0^2 \Omega_\mathrm{m0})/(8\pi Ga^3)$, the potential power spectrum is related to the density power spectrum by
\begin{equation}
\label{eq:potPower}
  P_\Phi(k)=\frac{9}{4}\frac{\Omega_\mathrm{m0}^2}{a^2}\frac{H_0^4}{k^4}P_\delta(k)\;.
\end{equation}
Since $P_\delta(k)\propto k$ for $k\ll k_0$ and $P_\delta(k)\propto k^{-3}$ for $k\gg k_0$, where $k_0$ is the comoving wave number of the perturbation mode entering the horizon at matter-radiation equality, we have $P_\Phi(k)\propto k^{-3}$ for $k\ll k_0$ and $P_\Phi(k)\propto k^{-7}$ for $k\gg k_0$.

Due to the steepness of the power spectrum, we have to introduce a cut-off wave number $k_\mathrm{min}$ when evaluating the spectral moments,
\begin{equation}
\label{eq:specMomPot}
\sigma_j^2=\int\limits_{k_\mathrm{min}}^{\infty}\frac{k^2\mathrm{d}k}{2\pi^2}P_\Phi(k)\hat{W}_R^2(k)\;.
\end{equation}
Thus, $k_\mathrm{min}$ defines a sharp high-pass filter in $k$-space. It has to be chosen properly to filter out large potential modes and therefore also large-scale potential gradients responsible for peculiar velocities of collapsed structures. In this way, this filter ensures that the gravitational potential of a structure is defined with respect to the large-scale potential value in its direct vicinity and that the constraint of a vanishing potential gradient is fulfilled for structures of all sizes. If they moved, they would not be counted when searching for potential minima and the number density derived in that way would be too small. We will discuss later how to find the proper $k_\mathrm{min}$.

\begin{figure}[t]
	\resizebox{\hsize}{!}{\includegraphics[width=\textwidth]{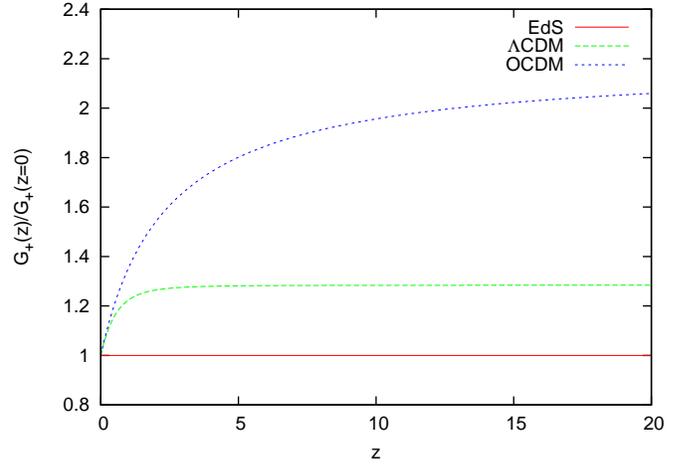}}
	\caption{Potential growth factor for three different cosmologies.}
	\label{fig:potGrowthFac}
\end{figure}

The evolution of the density power spectrum between the scale factors $a_1$ and $a_2$ is parametrised by the \emph{linear growth factor} $D_+(a)$ and the \emph{transfer function} $T(k,a)$,
\begin{equation}
  P_\delta\left(k,a_2\right)=\left[
    \frac{D_+\left(a_2\right)}{D_+\left(a_1\right)}
  \right]^2 T^2\left(k,a_2\right)P_\delta\left(k,a_1\right)\;.
\end{equation}
Since the transfer function $T(k,a)$ only changes for redshifts $z\ga 100$, we do not need to take it into account for the evolution of the power spectrum at $z<100$. Together with Eq.~(\ref{eq:potPower}), the evolution of the potential power spectrum is thus given by
\begin{equation}
\label{eq:evolutionPowerSpec}
  P_\Phi(k,a_2)=\left[\frac{G_+(a_2)}{G_+(a_1)}\right]^2P_\Phi(k,a_1)\;,
\end{equation}
which defines the \emph{potential growth factor} $G_+(a):=D_+(a)/a$. We normalise $G_+$ such that $G_+(a=1)=1$. Since $D_+(a)=a$ in an EdS universe, the potential growth factor stays constant, thus the potential power spectrum does not evolve with time in this case. This is not true for a $\Lambda$CDM and an OCDM model, yet the variation with time remains small. For cosmologies with $\Omega_\mathrm{m}\neq 1$, $\Omega_\Lambda\neq 0$ and negligible radiation density, the potential growth factor is accurately approximated by
\begin{eqnarray}
  G_+(a)&=&\frac{D_+(a)}{a}=\frac{5}{2}\Omega_\mathrm{m}(a)\Biggl\{
    \Omega_\mathrm{m}^{4/7}(a)-\Omega_\Lambda(a)\nonumber\\&+&
    \left[1+\frac{1}{2}\Omega_\mathrm{m}(a)\right]
    \left[1+\frac{1}{70}\Omega_\Lambda(a)\right]
  \Biggr\}^{-1}
\end{eqnarray}
\citep{Carroll1992}. Figure~\ref{fig:potGrowthFac} shows the evolution of $G_+$ with redshift. Obviously, the expression ``growth factor'' is somehow misleading because the potential power spectrum's amplitude is in fact \emph{decreasing} with time.

\subsection{Non-linear evolution}

Aiming at the number density of potential minima including non-linear evolution, we have to relate the potential from linear theory used so far, $\Phi_\mathrm{l}$, to the potential including non-linear evolution, $\Phi_\mathrm{nl}$. We shall use the spherical-collapse model to estimate the ratio $\Phi_\mathrm{nl}/\Phi_\mathrm{l}$.

\subsubsection{Gravitational potential in the centre of a homogeneous overdense sphere}

A spherical and homogeneous overdensity with density contrast $\delta$ and radius $R$ has a gravitational potential given by Poisson's equation,
\begin{equation}
  \frac{1}{r^2}\frac{\partial}{\partial r}\left(r^2\frac{\partial}{\partial r}\Phi\right)=
  4\pi G\bar{\rho} \theta(R-r)\;,
\end{equation}
where $\theta$ is \emph{Heaviside's step function} and $\bar{\rho}$ is the density inside the sphere acting as the source of the gravitational field. This equation holds in physical coordinates, while Eq.~(\ref{eq:relateSpectra}) uses comoving coordinates. Integrating twice with the boundary conditions $(\partial\Phi/\partial r) \rightarrow 0$ for $r\rightarrow 0$ and $\Phi\rightarrow 0$ for $r\rightarrow\infty$ yields
\begin{equation}
  \Phi(r)=\left\{\begin{array}{ll}
	-2\pi G\bar{\rho}R^2\left(1-\displaystyle{\frac{r^2}{3 R^2}}\right) &
	\quad\mbox{for}\quad r\leq R\;,\\
	-\displaystyle{\frac{4\pi G}{3}}\bar{\rho}\displaystyle{\frac{R^3}{r}} &
	\quad\mbox{else}\;,
  \end{array}\right.
\end{equation}
showing that the potential at the centre, $\Phi_0:=\Phi(r=0)$, is
\begin{equation}
\label{eq:centralPot}
  \Phi_0=-2\pi G\bar{\rho}R^2\;.
\end{equation}
Since the potential $\Phi(r)\propto 1-r^2/(3 R^2)$ inside a homogeneous sphere, it may appear appropriate to construct the low-pass filter from the window function
\begin{equation}
\label{eq:filter2real}
W_R(r)=\left(1-\frac{r^2}{3 R^2}\right)\left(\frac{16}{15}\pi R^3\right)^{-1}
\end{equation}
instead of a top-hat. It Fourier transforms to the filter function
\begin{equation}
\label{eq:filter2}
\hat{W}_R(k)=\frac{5\left[3\sin u-u\left(3+u^2\right)\cos u\right]}{2 u^5}\quad\mbox{with}\quad u=kR\;.
\end{equation}
Since the window function is more compact in real space than the top-hat, its Fourier transform is slightly broader in $k$-space and thus includes more Fourier modes.

This window function does not (and should not) reproduce the potential outside an isolated, homogeneous sphere, which drops $\propto r^{-1}$. This is no surprise because it drops to the potential of the homogeneous background within a finite radius, and the presence of a background potential signals the breakdown of Newtonian gravity in the cosmological context.

\subsubsection{Filtering radius}

The preceding consideration also provides a proper definition for the filtering radius of the window function $\hat{W}_R$ used for the calculation of the spectral moments, Eq.~(\ref{eq:topHat}). Poisson's equation for the perturbations in physical coordinates reads
\begin{equation}
\label{eq:poissonPhysical}
  \Delta_\mathrm{r}\Phi=4\pi G \bar{\rho}\;,
\end{equation}
with $\bar{\rho}=\rho_\mathrm{b}\delta$. Combining Eqs.~(\ref{eq:centralPot}) and (\ref{eq:poissonPhysical}), we see that the central potential is $\Phi_0=-\frac{1}{2}\Delta_\mathrm{r}\Phi R_\mathrm{r}^2$. We can use this relation to define a filtering radius in physical coordinates by
\begin{equation}
\label{eq:filteringRadius}
  R_\mathrm{r}:=\sqrt{\frac{-2\Phi}{\Delta_\mathrm{r}\Phi}}\;.
\end{equation}
This expression remains valid in comoving coordinates if we replace the Laplacian in physical coordinates by the Laplacian in comoving coordinates,
\begin{equation}
\label{eq:filteringRadiusII}
  R_\mathrm{com}=\frac{R_\mathrm{r}}{a}=
  \sqrt{\frac{-2\Phi}{a^2\Delta_\mathrm{r}\Phi}}=
  \sqrt{\frac{-2\Phi}{\Delta_\mathrm{com}\Phi}}\;.
\end{equation}

\subsubsection{Spherical collapse model}
\label{sec:SCM}

We rescale the scale factor $a$ and the radius of the overdense region $R$ by their values at turn-around, $a_\mathrm{ta}$ and $R_\mathrm{ta}$, respectively, defining the parameters
\begin{equation}
  x:=\frac{a}{a_\mathrm{ta}}\;,\quad y:=\frac{R}{R_\mathrm{ta}}\;.
\end{equation}
In addition, we introduce the dimensionless time $\tau$ and the overdensity at turn-around $\hat{\zeta}$ by
\begin{equation}
  \tau:=H_\mathrm{ta}t\;,\quad\rho_\mathrm{ta}=\hat{\zeta}\rho_\mathrm{b,ta}\;,
\end{equation}
with the Hubble parameter at turn-around $H_\mathrm{ta}$. In the following, we will sketch the most important steps to consider cosmologies like EdS, $\Lambda$CDM, and OCDM. Our consideration is based on \cite{Bartelmann2006}, but simplified because we ignore dynamical dark energy for now. This extends the work by \cite{Wang1998} towards cosmologies with non-vanishing curvature.

Spherical collapse is then described by the differential equations
\begin{eqnarray}
\label{eq:SCMI}
  x^\prime&=&\left[
  \frac{\Omega_\mathrm{m,ta}}{x}+\Omega_\mathrm{\Lambda,ta}x^2+
  (1-\Omega_\mathrm{m,ta}-\Omega_\mathrm{\Lambda,ta})
  \right]^{1/2}\;, \\
\label{eq:SCMII}
  y^{\prime\prime}&=&-\frac{\Omega_\mathrm{m,ta}\hat{\zeta}}{2y^2}+\Omega_\mathrm{\Lambda,ta}y\;,
\end{eqnarray}
with the matter and dark-energy densities at turn around, $\Omega_\mathrm{m,ta}$ and $\Omega_\mathrm{\Lambda,ta}$, respectively. Primes denote derivatives with respect to the dimensionless time $\tau$. Equations~(\ref{eq:SCMI}) and (\ref{eq:SCMII}) can be solved numerically using the boundary conditions $y=0$ at $x=0$ and $y^\prime=0$ at $x=1$, meaning that the overdensity starts with zero radius and reaches its maximal extent at turn-around. The requirement $y=1$ at turn-around then uniquely determines $\hat{\zeta}$.

The overdensity inside a halo with respect to the \emph{background density} at any time is $\Delta=(x/y)^3 \hat{\zeta}$. Assuming that the collapsing halo virialises at the collapse redshift $z_\mathrm{c}$, which corresponds to the normalised scale factor $x_\mathrm{c}$ when it would ideally collapse to zero radius, the virial overdensity is
\begin{equation}
\label{eq:defDelta}
  \Delta_\mathrm{v}=\hat{\zeta}\left(\frac{x_\mathrm{c}}{y_\mathrm{c}}\right)^3=
  \hat{\zeta}\left(\frac{R_\mathrm{ta}}{R_\mathrm{v}}\right)^3
  \left(\frac{1+z_\mathrm{ta}}{1+z_\mathrm{c}}\right)^3\;.
\end{equation}
The redshifts at collapse and turn-around, $z_\mathrm{c}$ and $z_\mathrm{ta}$, are related by $t(z_\mathrm{ta})=\frac{1}{2}t(z_\mathrm{c})$. Later, we shall need the overdensity with respect to the \emph{critical density} rather than the background density, $\tilde{\Delta}_\mathrm{v}=\Omega_\mathrm{m}(z_\mathrm{c})\Delta_\mathrm{v}$. The ratio between the radii at turn-around and at collapse is 2 in an EdS universe because of virialisation, but slightly different in more general cosmologies. Due to an additional contribution of dark energy to the potential, denoted by $\langle E_\mathrm{de}\rangle$, the virial theorem is modified to $\langle E_\mathrm{kin}\rangle=-\frac{1}{2}\langle E_\mathrm{pot}\rangle+\langle E_\mathrm{de}\rangle$. This leads to the following approximation for the ratio of the two radii \citep{Wang1998}:
\begin{eqnarray}
  y_\mathrm{c}&=&\frac{R_\mathrm{v}}{R_\mathrm{ta}}=
  \frac{1-\frac{1}{2}\eta_\nu}{2+\eta_\mathrm{t}-\frac{3}{2}\eta_\nu}\;, \nonumber \\
  \eta_\nu&=&\frac{2}{\hat{\zeta}}
  \frac{\Omega_\Lambda(z_\mathrm{c})}{\Omega_\mathrm{m}(z_\mathrm{c})}
  \left(\frac{1+z_\mathrm{c}}{1+z_\mathrm{ta}}\right)^3\;,
  \quad\eta_\mathrm{t}=
  \frac{2}{\hat{\zeta}}\frac{\Omega_\Lambda(z_\mathrm{ta})}{\Omega_\mathrm{m}(z_\mathrm{ta})}\;.
\end{eqnarray}
As the density perturbations evolve linearly at very early times, the density contrast expected from linear theory at collapse, $\delta_\mathrm{c}(x_\mathrm{c})$, is simply given by
\begin{equation}
  \delta_\mathrm{c}(x_\mathrm{c})=\lim \limits_{x \to 0}
  \left\{\frac{D_+(x_\mathrm{c})}{D_+(x)}\left[\Delta_\mathrm{v}(x)-1\right]\right\}\;.
\end{equation}
In an EdS universe, $\delta_\mathrm{c}=1.686$ and $\Delta_\mathrm{v}=178$, both independent of the collapse redshift.

\subsection{Ratio of linearly and non-linearly evolved potential}

We shall use the potential in the centre of a spherical and homogeneous overdensity, Eq.~(\ref{eq:centralPot}), to relate the linear and non-linear potential depths. During the following calculation, we will denote quantities at an initial scale factor $a_\mathrm{i}$ with the subscript `i' and quantities at the collapse scale factor $a_\mathrm{c}$ with the subscript `c'.

The potentials at the initial time and at collapse are
\begin{equation}
  \Phi_\mathrm{i}=-2\pi G\bar{\rho}_\mathrm{i}R_\mathrm{ta}^2
  y_\mathrm{i}^2\;,\quad
  \Phi_\mathrm{c}=-2\pi G\bar{\rho}_\mathrm{c}R_\mathrm{ta}^2 y_\mathrm{c}^2\;,
\end{equation}
respectively. Their ratio is
\begin{equation}
  \frac{\Phi_\mathrm{c}}{\Phi_\mathrm{i}}=
  \frac{\delta_\mathrm{v}}{\delta_\mathrm{i}}\left(\frac{y_\mathrm{c}}{x_\mathrm{c}}\right)^3
  \left(\frac{x_\mathrm{i}}{y_\mathrm{i}}\right)^3\frac{y_\mathrm{i}}{y_\mathrm{c}}\;,
\end{equation}
where $\delta_\mathrm{v}=\Delta_\mathrm{v}-1$. We have used here that the densities at the initial time and at collapse time are
$\bar{\rho}_\mathrm{i}=\rho_\mathrm{b0}a_\mathrm{i}^{-3}\delta_\mathrm{i}$ and $\bar{\rho}_\mathrm{c}=\rho_\mathrm{b0}a_\mathrm{c}^{-3}\delta_\mathrm{v}$,
respectively. Using Eq.~(\ref{eq:defDelta}), we can write $(y_c/x_c)^3=\hat{\zeta}/\Delta_\mathrm{v}$ and $y_\mathrm{i}\approx\hat{\zeta}^{1/3}x_\mathrm{i}$ since $\Delta_\mathrm{i}\approx 1$ for early times. This yields
\begin{equation}
\label{eq:nonLinEvolution}
\frac{\Phi_\mathrm{c}}{\Phi_\mathrm{i}}=\frac{\delta_\mathrm{v}}{\Delta_\mathrm{v}}\frac{\hat{\zeta}^{1/3}}{y_\mathrm{c}}\frac{x_\mathrm{i}}{\delta_\mathrm{i}}\approx\frac{\hat{\zeta}^{1/3}}{y_\mathrm{c}}\frac{x_\mathrm{i}}{\delta_\mathrm{i}}\;.
\end{equation}
We have neglected the difference between $\delta_\mathrm{v}$ and $\Delta_\mathrm{v}$ in the last step, which is a good approximation since $\Delta_\mathrm{v}=\mathcal{O}(10^2)$. The ratio $\delta_\mathrm{i}/x_\mathrm{i}=:C$ is given by
\begin{equation}
\label{eq:nonLinRatio}
  C=\frac{3}{5}\left[\hat{\zeta}^{1/3}\left(
    1+\frac{\Omega_\mathrm{\Lambda,ta}}{\hat{\zeta} \Omega_\mathrm{m,ta}}
    \right)+\frac{1-\Omega_\mathrm{m,ta}-\Omega_\mathrm{\Lambda,ta}}{\Omega_\mathrm{m,ta}}
  \right]
\end{equation}
\citep{Bartelmann2006}. Since Eqs.~(\ref{eq:nonLinEvolution},\ref{eq:nonLinRatio}) describe the non-linear evolution between scale factors $a_\mathrm{i}$ and $a_\mathrm{c}$, the linear and non-linear potential depths are related by
\begin{equation}
\label{eq:linNonLin}
  \frac{\Phi_\mathrm{nl}}{\Phi_\mathrm{l}}=
  \frac{\Phi_\mathrm{c}}{\Phi_\mathrm{i}}\frac{\Phi_\mathrm{i}}{\Phi_\mathrm{l,c}}=
  \frac{\hat{\zeta}^{1/3}}{y_\mathrm{c}}\frac{1}{C}\frac{G_+(a_\mathrm{i})}{G_+(a_\mathrm{c})}\;,
\end{equation}
where $\Phi_\mathrm{l,c}$ is the potential evolved linearly from $a_\mathrm{i}$ to $a_\mathrm{c}$. For an EdS universe, we have $\Phi_\mathrm{nl}/\Phi_\mathrm{l}=\frac{10}{3}$, independent of the collapse redshift because the potential growth factor $G_+(a)$ is constant and $y_\mathrm{c}=\frac{1}{2}$, but for more general cosmological models, it depends on the collapse redshift.

\begin{figure}[t]
	\resizebox{\hsize}{!}{\includegraphics[width=\textwidth]{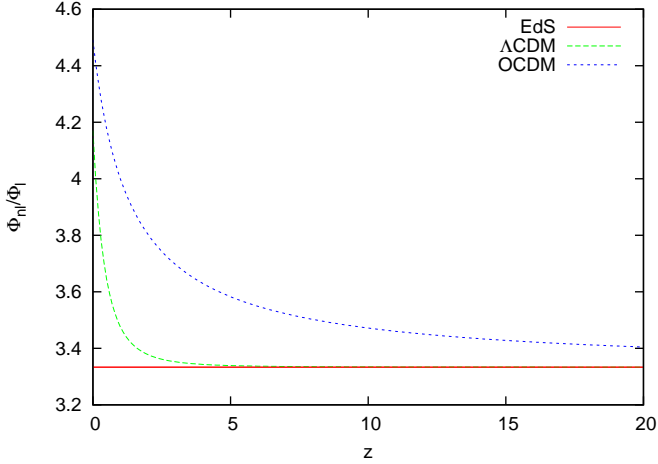}}
	\caption{The ratio of the potential depth that is expected at collapse redshift $z_\mathrm{c}$ from linear theory, $\Phi_\mathrm{l}$, and the one inferred from the SCM, $\Phi_\mathrm{nl}$.}
	\label{fig:linNonlinGrowth}
\end{figure}

We show $\Phi_\mathrm{nl}/\Phi_\mathrm{l}$ for three different cosmological models in Fig.~\ref{fig:linNonlinGrowth}. We arbitrarily choose the initial scale factor to be five times the scale factor at matter-radiation equality because this is early enough for the potential not to have begun developing since we are in the matter dominated era, and late enough for ignoring any evolution of the transfer function $T(k,a)$ with time. While the ratio Eq.~(\ref{eq:linNonLin}) reaches the constant $\frac{10}{3}$ expected in an EdS model relatively quickly in the $\Lambda$CDM case as the redshift increases, its evolution is much slower for OCDM. We point out that the difference between the linear and the non-linear potential evolution in the centre is by far not as large as for the density contrast, where it is of order $10^{3\ldots4}$.

\subsection{Counting only collapsed structures}

Since we are only interested in collapsed structures forming halos, we have to include this additional property in our calculations. Following \citet{Press1974}, this can be done by only taking structures into account whose linear density contrast exceeds the critical value $\delta_\mathrm{c}$. In our case, this translates to the criterion that the Laplacian $\Delta\Phi$ exceed a certain threshold $\Delta\Phi_\mathrm{c}$, which is, according to Poisson's equation, given by
\begin{equation}
  \Delta\Phi_\mathrm{c}(a)=\frac{3}{2}H_0^2\Omega_\mathrm{m0}\frac{\delta_\mathrm{c}(a)}{a}\;.
\end{equation}
In that way, we only count structures whose gravitational potential is ``curved enough'', so that the number density of potential minima belonging to collapsed structures is finally
\begin{equation}
\label{eq:intPotFunc}
  n(\Phi)=
  \int\limits_{\Delta\Phi_\mathrm{c}}^\infty\mathrm{d}(\Delta\Phi) \tilde{n}(\Phi,\Delta\Phi)\;,
\end{equation}
with $\tilde{n}(\Phi,\Delta\Phi)$ given by Eqs.~(\ref{eq:numDensPhiDelPhi}--\ref{eq:numDensPhiDelPhiLast}).

\section{Construction of the X-ray temperature function}

Combining the results of the previous sections, we are now able to derive a function describing the differential number density of structures as a function of their X-ray temperature based on the statistics of minima in the cosmic gravitational potential field. In order to count them properly, a high-pass filter is introduced that removes disturbing large modes. Assuming virial equilibrium, it is possible to relate the potential depths to X-ray temperatures using the virial theorem.

\subsection{Virial theorem}

We can relate the potential depth $\Phi_\mathrm{nl}$ to the X-ray temperature of a cluster via the virial theorem for Newtonian gravity, which relates the ensemble-averaged kinetic and potential energies, $\langle E_\mathrm{kin}\rangle$ and $\langle E_\mathrm{pot}\rangle$, by $\langle E_\mathrm{kin}\rangle=-\frac{1}{2}\langle E_\mathrm{pot}\rangle$. The kinetic energy is connected to a temperature $T$ by $\langle E_\mathrm{kin}\rangle=\frac{3}{2}k_\mathrm{B}T$, where $k_\mathrm{B}$ is Boltzmann's constant. The potential energy is $m\Phi$ with a proper mass $m$. Assuming that the intracluster medium is fully ionised, we know that $m=\mu m_\mathrm{p}$ with $\mu=0.59$. Particles near the cluster core feel the potential $\Phi(r)\approx\Phi_0$, thus the virial theorem reads\footnote{Strictly speaking, this is only valid for a universe with $\Omega_\Lambda=0$. Due to the presence of dark energy, an additional potential arises \citep{Wang1998} whose contribution is small (e.g. the virial radius only changes about few percents, see Sect.~\ref{sec:SCM}), so that we neglect it in our further calculations. Whether dark energy contributes at all to the process of virialisation is still an open question, see e.g. the discussion by \citet{Bartelmann2006}.}
\begin{equation}
\label{eq:virTheorem}
  -\mu m_\mathrm{p}\Phi_0=3k_\mathrm{B}T\;.
\end{equation}

Since $\Phi_0$ can be regarded as the non-linear depth of a potential minimum, we can replace $\Phi_\mathrm{nl}$ by $\Phi_0=-3 k_\mathrm{B} T/(\mu m_\mathrm{p})$. Given a particular X-ray temperature, we calculate the corresponding linear potential depth by relating the temperature to the non-linear potential using Eq.~(\ref{eq:virTheorem}), and the latter to the linear potential using Eq.~(\ref{eq:linNonLin}), arriving at
\begin{equation}
\label{eq:transPhiT}
  \Phi_\mathrm{l}=-\frac{3y_\mathrm{c}}{\hat{\zeta}^{1/3}}
  \frac{C}{\mu m_\mathrm{p}}\frac{G_+(z)}{G_+(z_\mathrm{i})}k_\mathrm{B} T\;.
\end{equation}

\subsection{Evaluation steps}

The X-ray temperature function is implicitly determined by Eqs.~(\ref{eq:numDensPhiDelPhi}--\ref{eq:numDensPhiDelPhiLast}). Assuming a temperature $T$, the linear potential depth $\Phi_\mathrm{l}$ is found from Eq.~(\ref{eq:transPhiT}). Then, Eq.~(\ref{eq:numDensPhiDelPhi}) needs to be integrated over $\Delta\Phi$ from $\Delta\Phi_\mathrm{c}$ to infinity. Since the smoothing radius (\ref{eq:filteringRadiusII}) depends on the integration variable $\Delta\Phi$, each step in the integration requires updating the spectral moments, cf.~Eq.~(\ref{eq:specMoments}).

In practice, the temperature interval for which the X-ray temperature function is to be calculated is divided into a reasonable number of bins large enough that the shape of the temperature function can be inferred from interpolating between them. At first, each X-ray temperature $T_i$ corresponding to one particular bin has to be related to a linear potential depth $\Phi_{\mathrm{l},i}$ using Eq.~(\ref{eq:transPhiT}). We also have to take care of the Jacobian determinant of the transformation because $n(T) \mathrm{d}T$ must be replaced by $n(\Phi_\mathrm{l}) (\mathrm{d}T/\mathrm{d}\Phi_\mathrm{l}) \mathrm{d}\Phi_\mathrm{l}$. Having related both quantities, it is possible to evaluate Eqs.~(\ref{eq:numDensPhiDelPhi}--\ref{eq:numDensPhiDelPhiLast}).

In order to calculate the spectral moments $\sigma_0$, $\sigma_1$, and $\sigma_2$, we also have to choose a reasonable amount of bins for the Laplacian of the potential, which we will denote with $\Delta\Phi_j$, since the spectral moments are functions of both $\Phi_\mathrm{l}$ and $\Delta\Phi$ through the filtering radius entering via the Fourier transform of the window function, Eq.~(\ref{eq:topHat}). Additionally, the cut-off wave vector $k_\mathrm{min}$ defining a sharp high-pass filter in $k$-space is also a function of both quantities. A detailed discussion how to find the proper $k_\mathrm{min}$ for a given temperature and Laplacian of the gravitational potential is presented in Sect.~\ref{sec:highPass}. Thus, we have to evaluate both the filtering radius and $k_\mathrm{min}$ for each pair $(\Phi_{\mathrm{l},i},\Delta\Phi_j)$ using Eq.~(\ref{eq:filteringRadiusII}) and starting for a given $\Phi_{\mathrm{l},i}$ with $\Delta\Phi_0=\Delta\Phi_\mathrm{c}$.

Inserting the spectral moments into Eqs.~(\ref{eq:numDensPhiDelPhi}--\ref{eq:numDensPhiDelPhiLast}) yields the number density of minima per potential interval $\mathrm{d}\Phi_\mathrm{l}$ and per interval of the Laplacian $\mathrm{d}(\Delta\Phi)$ for the specific parameter pair $(\Phi_{\mathrm{l},i},\Delta\Phi_\mathrm{c})$. Since $\Delta\Phi_\mathrm{c}=\mathcal{O}(10^4\ \mbox{km}^2\ \mbox{s}^{-2}\ \mbox{Mpc}^{-2}\ h^2)$, we choose the next step to be $\Delta\Phi_\mathrm{1}=\Delta\Phi_\mathrm{c}+\delta(\Delta\Phi)$ with $\delta(\Delta\Phi)=10^2\ \mbox{km}^2\ \mbox{s}^{-2}\ \mbox{Mpc}^{-2}\ h^2$ and calculate now $R$ and the spectral moments for the pair $(\Phi_{\mathrm{l},i},\Delta\Phi_1)$. We continue with $\Delta\Phi_j=\Delta\Phi_{j-1}+\delta(\Delta\Phi)$ for a given $\Phi_{\mathrm{l},i}$ until we fulfill the following convergence criterion.

We approximate the integral in Eq.~(\ref{eq:intPotFunc}) by the trapezium rule. Hence, for a specific potential value $\Phi_{\mathrm{l},i}$, the number density $n(\Phi_{\mathrm{l},i})$ is evaluated numerically as
\begin{equation}
  n(\Phi_{\mathrm{l},i})\approx\frac{\delta(\Delta\Phi)}{2}\sum_{j=1}^N\left[
    \tilde{n}(\Phi_{\mathrm{l},i},\Delta\Phi_{j-1})+\tilde{n}(\Phi_{\mathrm{l},i},\Delta\Phi_j)
  \right]\;.
\end{equation}
This summation is stopped at an index $N$ chosen as the first index for which the relative contribution to $n(\Phi_{\mathrm{l},i})$ is smaller than $10^{-6}$. This is a proper break condition since $\tilde{n}(\Phi_\mathrm{l},\Delta\Phi)$ tends rapidly towards zero for $\Delta\Phi\to\infty$.

Having evaluated $n(T_i)$ for each $i$, we have an appropriate approximation of the X-ray temperature function's shape in the chosen temperature interval.

\subsection{High-pass filtering}
\label{sec:highPass}

As mentioned before, we need a proper high-pass filter in order to remove disturbing large-scale potential modes since we want the potential associated with a collapsed structure to be defined with respect to the large-scale potential level in its direct vicinity, and we want the structure to have no peculiar motion so that the constraint $\vec{\eta}=\vec{0}$ is applicable. The natural filter choice is a sharp cut-off in $k$-space since this will effectively remove both large-scale potential modes and potential gradients.

Let again $R$ be the top-hat filter radius of Eq.~(\ref{eq:topHat}) and $R_\mathrm{hp}$ the filter radius related to a sharp cut-off wave number $k_\mathrm{hp}$ in $k$-space by $k_\mathrm{hp}=2\pi/R_\mathrm{hp}$. The ratio of both quantities defines $\alpha$,
\begin{equation}
R_\mathrm{hp}=\alpha R\;.
\end{equation}
Since $R$ and $R_\mathrm{hp}$ define a low- and a high-pass filter, respectively, one can expect $\alpha>1$.

\begin{figure}[t]
\begin{center}
\includegraphics[width=0.49\hsize]{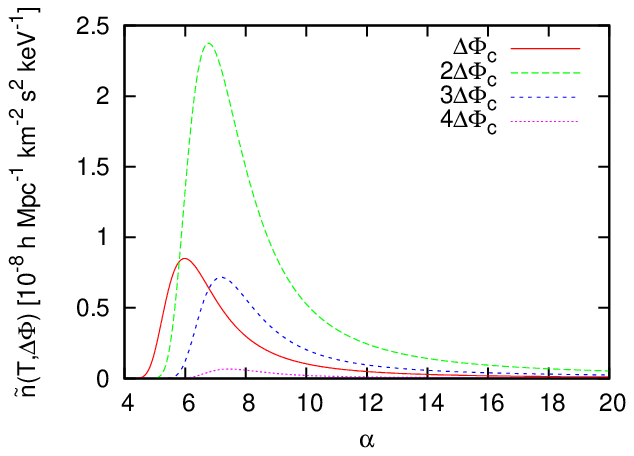}
\includegraphics[width=0.49\hsize]{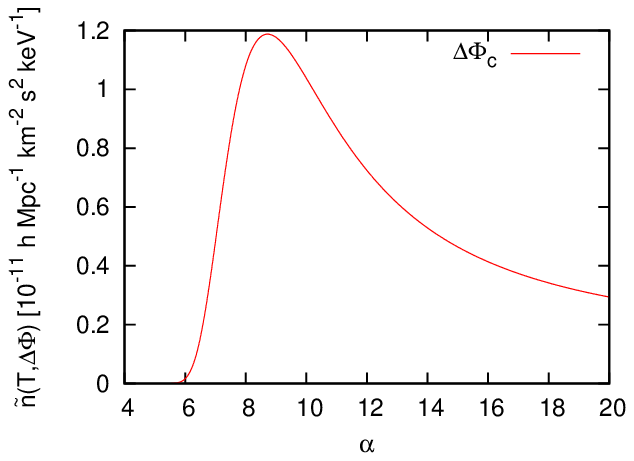}\hfill
\includegraphics[width=0.49\hsize]{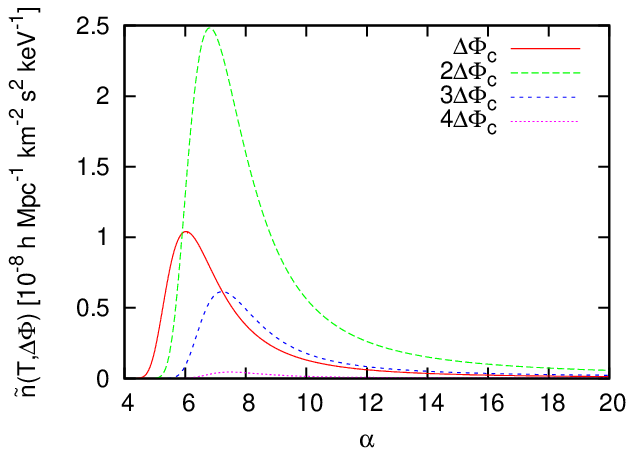}
\includegraphics[width=0.49\hsize]{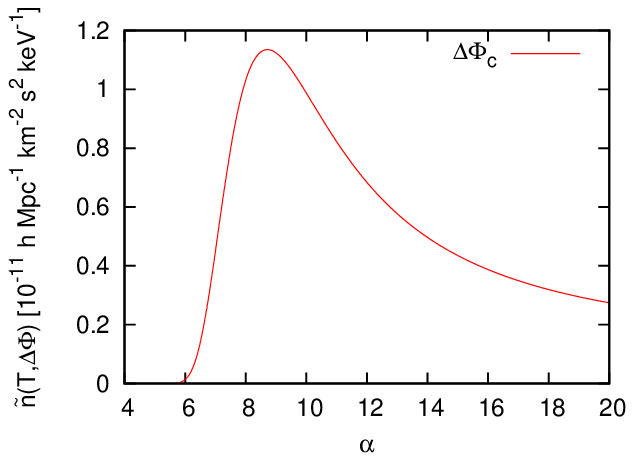}\hfill
\includegraphics[width=0.49\hsize]{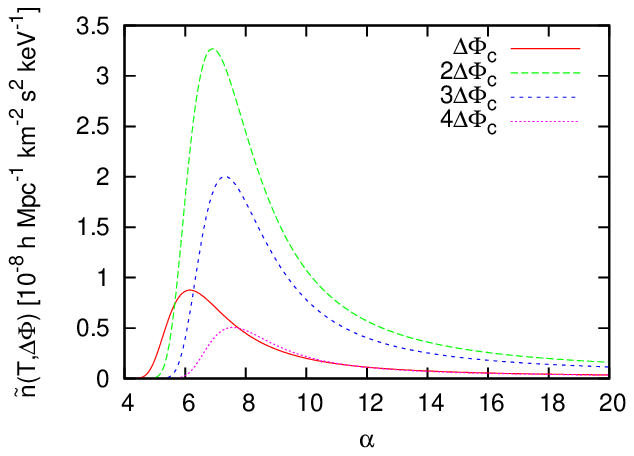}
\includegraphics[width=0.49\hsize]{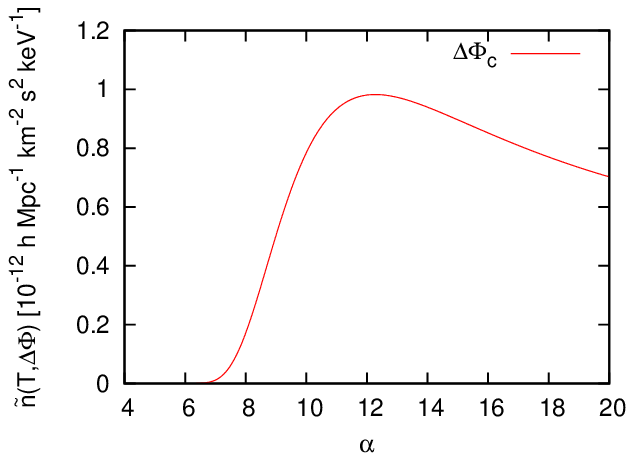}\hfill
\end{center}
\caption{Differential number density $\tilde{n}(T,\Delta\Phi)$ as function of the parameter $\alpha$ for three different cosmologies and two temperatures at $z=0$. \textit{Upper row:} $\Lambda$CDM. \textit{Central row:} OCDM. \textit{Lower row:} EdS. \textit{Left panels:} $T=1$~keV. \textit{Right panels:} $T=10$~keV. The values for $\Delta\Phi$ were chosen to be multiples of the critical Laplacian $\Delta\Phi_\mathrm{c}$ and are given in the plots. For $T=10$~keV we only plot the number density for $\Delta\Phi=\Delta\Phi_\mathrm{c}$ since it is a very steep function of $\Delta\Phi$ and it is too small for larger multiples of $\Delta\Phi_\mathrm{c}$ to be seen.} 
\label{fig:maxima}
\end{figure}

It is now a legitimate question how $\alpha$ should be chosen for calculating the correct number density of objects having a particular potential depth. We argue that a natural choice exists. In Fig.~\ref{fig:maxima}, we plot $\tilde{n}(T,\Delta\Phi)$ as a function of $\alpha$. For each pair $(T_i,\Delta\Phi_k)$, the number density peaks at some value $\alpha_\mathrm{max}$, which is a function of both the temperature and the Laplacian of the potential. It increases with both increasing Laplacian and increasing temperature. For low temperatures, the maximum is rather sharp, but broadens as the temperature increases. This is most pronounced for the EdS universe. This behaviour can be understood as follows.

On the one hand, decreasing the radius of the high-pass filter starting from a value much larger than the typical size of the object excludes more and more modes on decreasing scales. In this way, potential \emph{gradients} are removed which would cause a non-zero peculiar velocity and thus a deviation from the constraint $\vec{\eta}=\vec{0}$. Since more and more objects are put to rest, the number density of objects with vanishing potential gradient increases. Figure~\ref{fig:maxima} shows that the increase of the number density is less steep for large (or hot) than for small (or cool) objects. This reflects the fact that that massive, hot objects, e.g.~with $T=10$~keV, are more likely located at potential minima, thus removing large-scale modes has less effect on their number counts.

On the other hand, increasing the radius of the high-pass filter starting from a value much smaller than the typical size of the relevant object adds more and more modes. While the window between low- and high-pass filtering is too small, modes relevant for the structures considered are filtered out and the halo number density remains approximately zero. Once modes are included that compose the structures, the halo number density steeply rises as $\alpha$ is increased.

At a certain $\alpha_\mathrm{max}$, the number density of objects reaches a maximum where both effects are balanced. Then, all modes relevant for structures of size $R=\sqrt{-2\Phi_i/\Delta\Phi_j}$ are included, but larger modes are excluded which would create a non-vanishing potential gradient. Thus, $\alpha_\mathrm{max}$ can be used to define $k_\mathrm{min}$ for each pair $(T_i,\Delta\Phi_j)$ individually by
\begin{equation}
\label{eq:kMin}
k_\mathrm{min}=\frac{2\pi}{\alpha_\mathrm{max}R}\;.
\end{equation}
This definition of $k_\mathrm{min}$ has to be used when evaluating the spectral moments with Eq.~(\ref{eq:specMomPot}). 

In appendix~\ref{ap:kmin}, we present an alternative way to define a physically reasonable cut-off wave vector for the evaluation of the spectral moments. It turns out, however, that the number density of objects with a low X-ray temperature function is highly underestimated in this way.

\subsection{Inferring the temperature function from the Press-Schechter mass function}

In order to compare the X-ray temperature function that we have derived from the statistics of gravitational potential perturbations to the classical Press-Schechter theory, we need a proper, albeit idealised, and consistent mass-temperature relation. Since we used the virial theorem, Eq.~(\ref{eq:virTheorem}), to relate the temperature to the potential, we will start at the same point to relate the temperature to a mass. Note that this has nothing to do with an assumption on real clusters, but merely serves the purpose of a theoretical cross-comparison between the mass-based Press-Schechter approach and our direct derivation of the temperature function.

We saw earlier that for a spherical and homogeneous overdensity the potential depth in the centre is $\Phi_0=-2\pi G \bar{\rho} R^2$, where $\bar{\rho}=\delta_\mathrm{v}\rho_\mathrm{b}$ is the constant density inside the \emph{perturbation}. Nonetheless, we shall replace $\delta_\mathrm{v}$ by the virial overdensity $\Delta_\mathrm{v}$, which is a good approximation because $\delta_\mathrm{v}=\Delta_\mathrm{v}-1$ and $\Delta_\mathrm{v}=\mathcal{O}(10^2)$.

The mass of the overdensity is $M=\frac{4}{3}\pi\rho R^3$, where $\rho$ is the total density inside the sphere. It is related to the background density by $\rho=\Delta_\mathrm{v}\rho_\mathrm{b}$. According to the previous statements, we can identify $\bar{\rho}$ and $\rho$, thus $\bar{\rho}\approx\rho$. Combining the equations for the potential and the mass and using the virial theorem yields the temperature-mass relation
\begin{equation}
\label{eq:tempM}
  k_\mathrm{B} T=\left(\frac{\pi\rho}{6}\right)^{1/3} G \mu m_\mathrm{p}  M^{2/3}\;.
\end{equation}
The density inside the cluster $\rho$ is related to the background density by $\rho(z)=\rho_\mathrm{cr0}\Omega_\mathrm{m0}\Delta_\mathrm{v}(z) (1+z)^3$. In our further calculations, however, we will use the virial overdensity $\tilde{\Delta}_\mathrm{v}$, which was defined with respect to the \emph{critical} density instead of the \emph{background} density. This yields
\begin{equation}
  \rho(z)=\rho_\mathrm{cr0}
  \frac{\Omega_\mathrm{m0}}{\Omega_\mathrm{m}(z)}\tilde{\Delta}_\mathrm{v}(z)(1+z)^3\;.
\end{equation}
Inserting the previous equation into Eq.~(\ref{eq:tempM}), normalising to $M=10^{15}M_\odot$ and $\tilde{\Delta}_\mathrm{v}=178$, and combining all other quantities including $H_0=100\,h\,\mathrm{km\,s^{-1}\,Mpc^{-1}}$ into one normalisation factor, we find
\begin{equation}
\label{eq:massTempFinal}
  k_\mathrm{B} T=7.83\ \mbox{keV}\left(
    \frac{\Omega_\mathrm{m0}}{\Omega_\mathrm{m}(z)}\frac{\tilde{\Delta}_\mathrm{v}(z)}{178}
  \right)^{1/3} \left(\frac{M}{10^{15}M_\odot h^{-1}}\right)^{2/3}(1+z)\;.
\end{equation}
This is almost the same relation as given by \cite{Eke1996}. The only difference is the normalisation factor. While we used a spherical and homogeneous overdensity for the calculation in order to be consistent with the derivation of the potential function, they derived a mass temperature-relation for an isothermal sphere, which results in a different normalisation.

We shall use this relation to convert the classical Press-Schechter mass function into a temperature function. Again, we must not forget to account for the Jacobian when transforming from mass to temperature. In contrast to the transformation of the potential to the temperature, the Jacobian of the mass-temperature transformation depends on temperature since both quantities are non-linearly related.

Finally, we also use Eq.~(\ref{eq:massTempFinal}) to convert the mass functions derived by \citet{Sheth1999} and \citet{Jenkins2001} to X-ray temperature functions knowing that these are fits to numerical simulations including effects of ellipsoidal collapse, while the derivation of our mass-temperature relation is based on spherical collapse. Yet, it should yield qualitative information on the importance of ellipsoidal collapse in our potential approach, since structures in the potential field are smoother than structures in the density field.

\section{Results}

\begin{figure*}[t]
\begin{center}
\includegraphics[width=0.33\hsize]{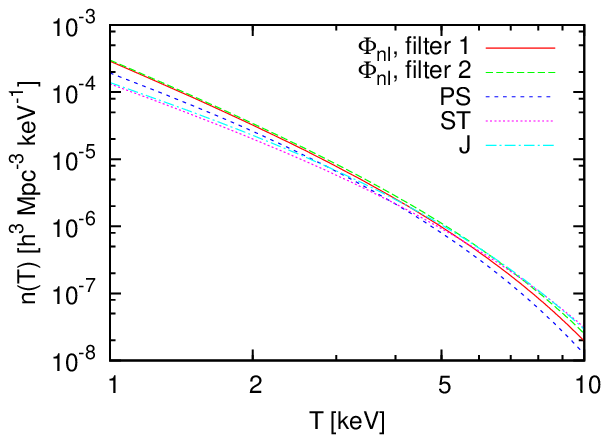}
\includegraphics[width=0.33\hsize]{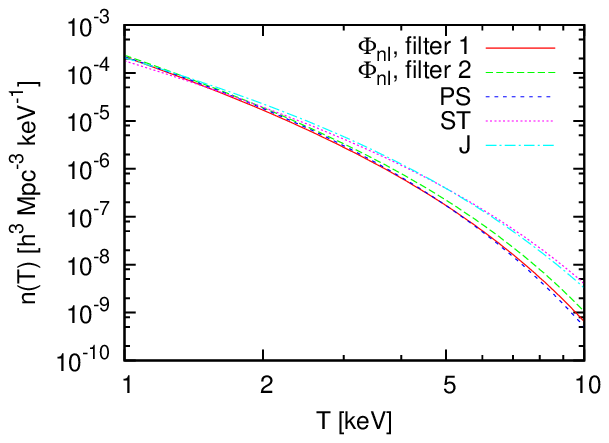}
\includegraphics[width=0.33\hsize]{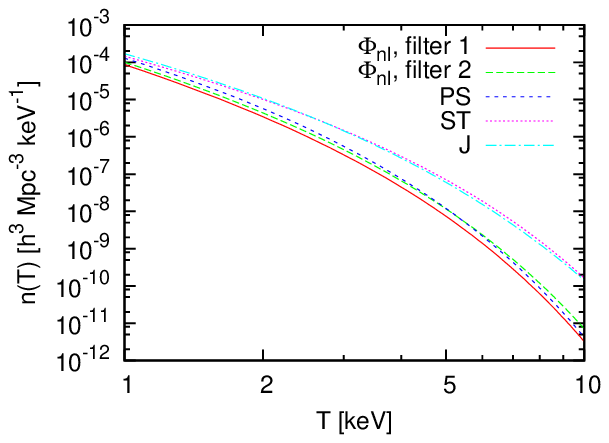}\hfill
\includegraphics[width=0.33\hsize]{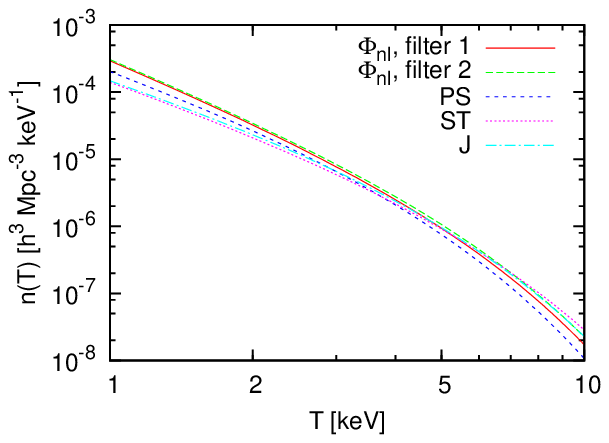}
\includegraphics[width=0.33\hsize]{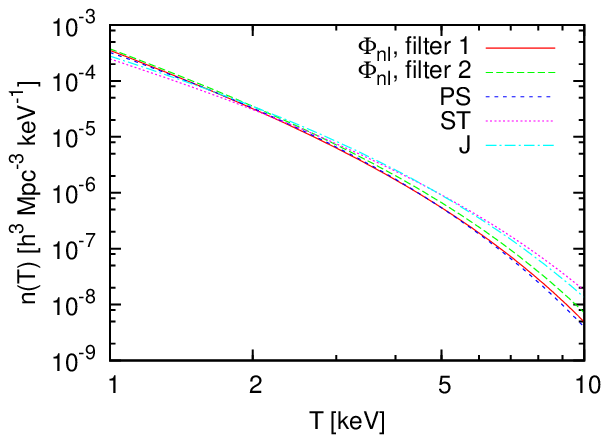}
\includegraphics[width=0.33\hsize]{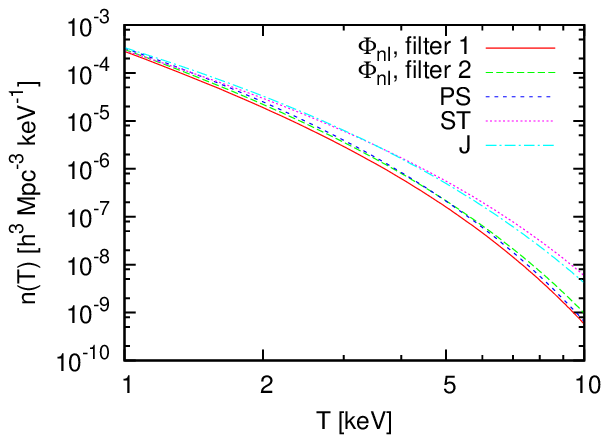}\hfill
\includegraphics[width=0.33\hsize]{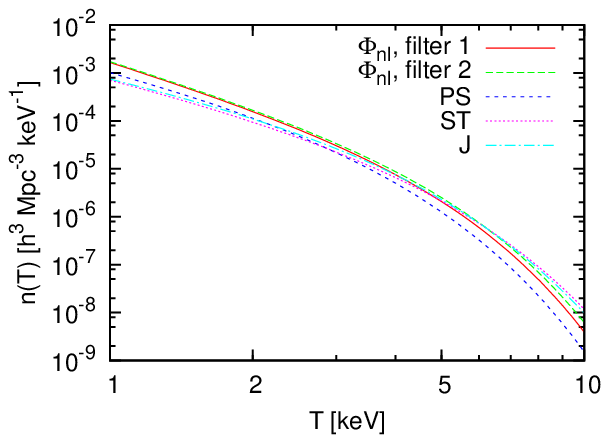}
\includegraphics[width=0.33\hsize]{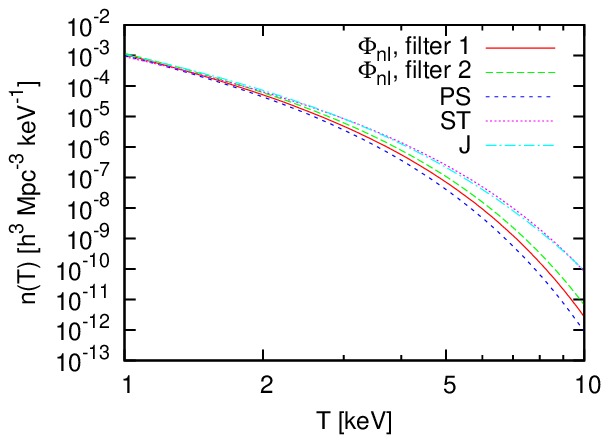}
\includegraphics[width=0.33\hsize]{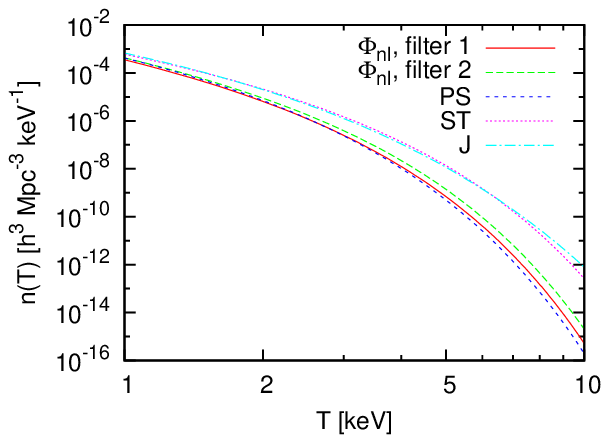}\hfill
\end{center}
\caption{Comparison of the X-ray temperature functions derived from the statistics of gravitational potential perturbations ($\Phi_\mathrm{nl}$) and the mass functions by Press~\&~Schechter (PS), Sheth~\&~Tormen (ST), and Jenkins~et~al. (J) for three different cosmologies and three redshifts. The filters 1 and 2 are defined by Eqs.~(\ref{eq:topHat}) and (\ref{eq:filter2}), respectively. \textit{Upper row:} $\Lambda$CDM. \textit{Central row:} OCDM. \textit{Lower row:} EdS. \textit{Left panels:} $z=0$. \textit{Central panels:} $z=1$. \textit{Right panels:} $z=2$. We want to emphasise that the y-axes are not scaled equally in order to compare the temperature functions more appropriately.}
\label{fig:tempFunc}
\end{figure*}

In this section, we present the X-ray temperature functions for the $\Lambda$CDM, OCDM, and EdS cosmologies calculated from the statistics of gravitational potential perturbations and compare them to temperature functions derived from three well-known mass functions. The results are shown in Fig.~\ref{fig:tempFunc}, where we compare the X-ray temperature function inferred from the potential using Eq.~(\ref{eq:transPhiT}) to the X-ray temperature function calculated from the Press-Schechter, the Sheth-Tormen and the Jenkins~et~al.\ mass functions using Eq.~(\ref{eq:massTempFinal}) for our three cosmological models and three redshifts.

The X-ray temperature function inferred from the statistics of fluctuations in the gravitational potential matches the temperature function inferred from the Press-Schechter mass function quite well for all three cosmological models and redshifts shown. As expected, the filter modelled after the internal potential profile of a homogeneous sphere (filter 2) yields slightly larger number densities especially at the high-temperature end because it is wider in $k$-space than the top-hat filter (filter 1). The temperature function based on our novel approach is in good agreement with the classical Press-Schechter approach with an only slightly different amplitude depending on redshift. These differences, however, may be irrelevant because of the idealising assumptions entering both approaches.

A comparison with temperature functions inferred from mass functions including elliptical collapse like the Sheth-Tormen mass function shows that the deviations increase substantially, especially for high redshifts. This suggests that ellipsoidal collapse should also explicitly be included in the potential approach. However, we emphasise again that this comparison is only qualitative since the mass-temperature relation used is based on spherical and homogeneous objects so that the actual difference between the different approaches may be even smaller.

\section{Discussion and Conclusions}

In this paper, we developed a novel approach to a theoretical derivation of an X-ray temperature function that does not depend on global, unobservable cluster quantities, but merely on the depth of the gravitational potential, which is a locally defined quantity directly related to observables. Using the statistical properties of a Gaussian random field, we were able to derive a distribution for the minima in the cosmic gravitational potential. Counting only those potential minima which are ``curved enough'', in the sense that their Laplacian exceeds a critical value, it was possible to semi-analytically compute a potential function belonging to collapsed structures. The critical Laplacian needed to distinguish collapsed structures from non-collapsed structures could be related to the critical density contrast $\delta_\mathrm{c}$ of the spherical-collapse model, which also plays an important role in the Press-Schechter formalism.

We also managed to calculate the influence of non-linear structure formation on the potential by referring again to the spherical-collapse model, which allows the ratio of the linearly and non-linearly evolved potential depths to be computed. Both the linear and the non-linear evolution of the potential are much slower compared to the evolution of the matter density. The ratio between the non-linearly and linearly evolved potential depths, for example, is $\sim30$ times smaller than the ratio $\Delta_\mathrm{v}/\delta_{c}$.

One crucial ingredient to our approach is a proper high-pass filter that removes large-scale potential modes and gradients resulting from them. We find that choosing the filter scale such as to maximise the number density of objects yields good agreement between the Press-Schechter approach based on an idealised mass-temperature relation, and our direct derivation of the temperature function. This criterion thus provides the foundation for using number counts of galaxy clusters in cosmology without invoking global quantities like cluster masses. Although we believe that our method is essentially ready to be tested with simulations before applying it to observational data, it leaves room for improvements in the theoretical description, as follows.

Based on this work, it should be examined if elliptical collapse can improve the results compared to fully non-linear $N$-body simulations than spherical collapse. First, we need to find out if the formalism for the elliptical collapse developed by \citet{Sheth1999} and \citet{Sheth2001} can be adapted for the potential calculation or if another approach must be found. Second, it will be interesting to see whether this will influence the results significantly, as suggested by Fig.~\ref{fig:tempFunc}. This is not as obvious as for density perturbations because potential fluctuations are much less asymmetric, and their non-linear evolution is much less pronounced than for density perturbations.

We calculated the X-ray temperature function from the potential function simply by applying the virial theorem relating the kinetic energy to the potential energy. It is straightforward to relate a temperature to a potential depth in this way. But there are two issues that should be examined in more detail. The first point is the influence of dark energy on the virialisation process. We have already mentioned that it might not affect virialisation at all, but if it does, it should be relatively small since the variation of the virial radius due to the additional potential originating from the dark energy is at the level of few per cent. A better understanding could strengthen the assumption that it can be neglected when relating the potential to a temperature for models including dark energy like $\Lambda$CDM. The second point is that we used the potential at the minimum of the potential well in the virial theorem. But since the virial theorem has to be applied to an averaged potential $\langle\Phi\rangle$ instead of the potential in the minimum $\Phi_0$, this should result in an additional correction factor when using $\Phi_0$ instead. A closer examination of this factor, especially its magnitude and its dependence on the potential depth, needs to be carried out. Besides, real clusters have lower temperatures in their centres due to cooling, implying that the application to observations requires calibration.

On the whole, our results are very promising and suggest to continue following this approach, which should allow a direct comparison of the cluster population with cosmological predictions based on observable, local cluster quantities. Although we only presented results for $\Lambda$CDM, OCDM, and EdS, it is straightforward to extend our computation of the X-ray temperature function to more elaborate cosmological models including e.g.~quintessence or early dark energy.

\acknowledgements{CA wants to thank M.~Maturi for providing realisations of Gaussian random fields and M.~Ecker, J.-C.~Waizmann, and F.~Pace for clarifying discussions. CA is supported by the SFB~439 of the Deutsche Forschungsgemeinschaft, the Heidelberg Graduate School of Fundamental Physics and the IMPRS for Astronomy \& Cosmic Physics at the University of Heidelberg.}

\bibliography{9562bib}

\begin{appendix}
\section{An alternative way to determine the cut-off wave number}
\label{ap:kmin}

In this appendix, we present an alternative approach to determine a physically reasonable definition of $k_\mathrm{min}$. Although it does not give the correct number density for smaller objects with a low X-ray temperature, the definition presented here may become important for future work based on the potential perturbation approach.

\subsection{Definition}

An alternative appropriate choice for $k_\mathrm{min}$ could be the redshift-dependent \emph{particle horizon} $r_\mathrm{hor}$, taking into account that light could have travelled only a finite comoving distance between the Big Bang and redshift $z$. Consequently, we must only consider modes of the gravitational potential already inside the horizon. Thus,
\begin{equation}
\label{eq:minWaveVector}
  k_\mathrm{min}(z)=\frac{\pi}{r_\mathrm{hor}(z)}=
  \frac{\pi H_0}{c}\left[
    \lim\limits_{a_1\rightarrow 0}
    \int\limits_{a_1}^{a(z)}\frac{\mathrm{d}a^\prime}{{a^\prime}^2E(a^\prime)}
  \right]^{-1}\;,
\end{equation}
where $E(a^\prime)$ is the \emph{expansion rate} of the Universe evaluated at the scale factor $a^\prime$.

In principle, signal retardation should also be taken into account. Considering an arbitrary point $\vec{x}_0$ at time $t_0$, only modes lying inside its past light cone can have influenced it. Retardation has to be included ``by hand'' because we are using Newtonian gravity. This gives rise to an additional factor when calculating the power spectrum's amplitude because we must evaluate the amplitude of a mode with wave length $\lambda$ not at time $t_0$, but at the earlier time $t_0-\Delta t=t_0-\lambda/(2c)$.

We can compute the corresponding scale factors at $t_0-\Delta t$ as follows. We must have $\lambda=2\pi/k=2D_\mathrm{com}(z,z_k)$, where $z$ and $z_k$ are the redshifts corresponding to times $t_0$ and $t_0-\Delta t$, respectively, and $D_\mathrm{com}(z, z_k)$ is the comoving distance between both redshifts. Thus, we have to find a scale factor $a_k=1/(1+z_k)$ for each mode $k$ such that
\begin{equation}
  k=\frac{\pi}{D_\mathrm{com}(a,a_k)}=
  \frac{\pi H_0}{c}\left[\int\limits_{a_k(z_k)}^{a(z)} \frac{\mathrm{d}a^\prime}{{a^\prime}^2 E(a^\prime)}\right]^{-1}
\end{equation}
holds. This is consistent with Eq.~(\ref{eq:minWaveVector}) because the wave number $k$ approaches $k_\mathrm{min}$ for $a_k\rightarrow0$. The influence on the power spectrum's amplitude results in an additional factor $G_+^2(a_k)/G_+^2(a)$ entering Eq.~(\ref{eq:evolutionPowerSpec}). Since $P_\Phi$ does not evolve with time in an EdS universe, it has no effect in this case. Additionally, it turns out that its contribution is quite small for both the $\Lambda$CDM and the OCDM model, where it only affects the power spectrum's amplitude for a relatively small amount of wave numbers. Hence, it only negligibly affects the computation of the spectral moments and can usually be ignored.

\subsection{Results for the X-ray temperature function}

\begin{figure}[t]
\begin{center}
\includegraphics[width=0.49\hsize]{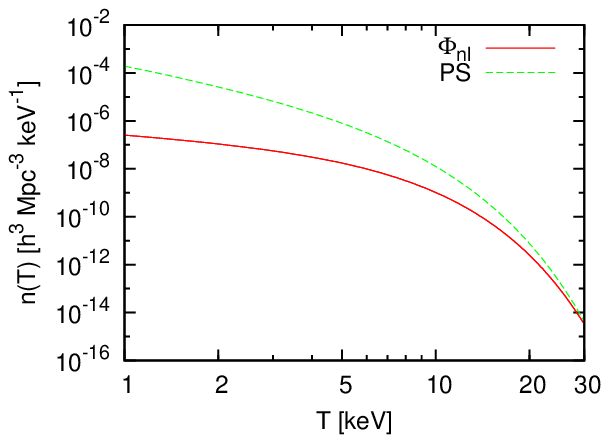}
\includegraphics[width=0.49\hsize]{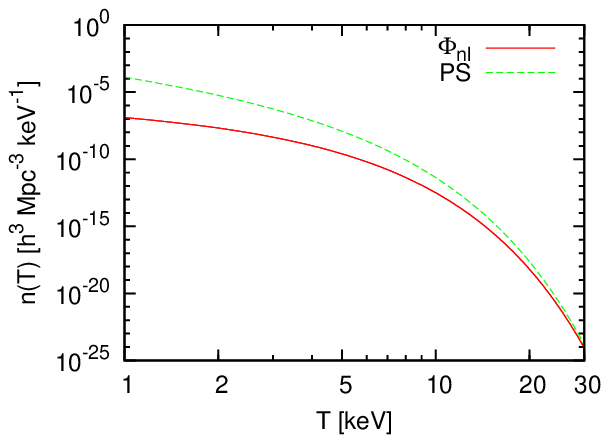}\hfill
\includegraphics[width=0.49\hsize]{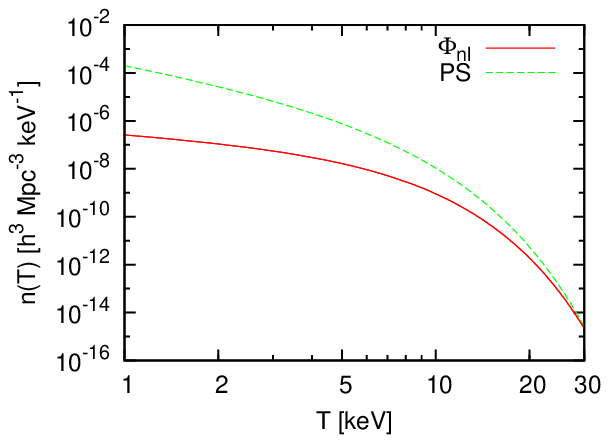}
\includegraphics[width=0.49\hsize]{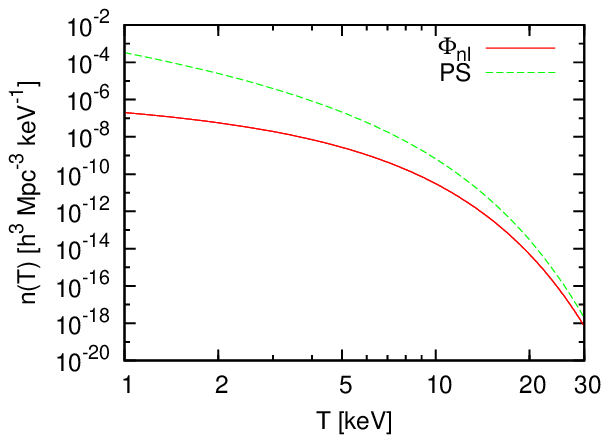}\hfill
\includegraphics[width=0.49\hsize]{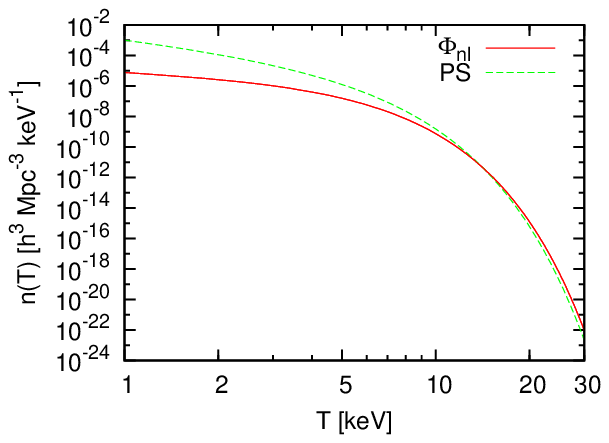}
\includegraphics[width=0.49\hsize]{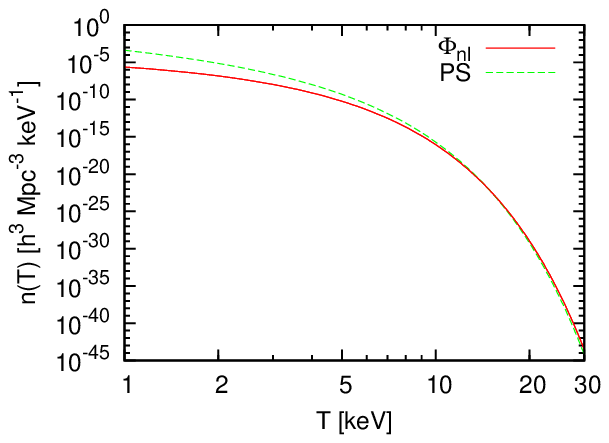}\hfill
\end{center}
\caption{Comparison of the X-ray temperature function derived from the statistics of gravitational potential perturbations ($\Phi_\mathrm{nl}$) using the alternative definition of $k_\mathrm{min}$ from Eq.~(\ref{eq:minWaveVector}) with the classical Press-Schechter approach (PS) for three cosmologies. \textit{Upper row:} $\Lambda$CDM. \textit{Central row:} OCDM. \textit{Lower row:} EdS. \textit{Left panels:} $z=0$. \textit{Right panels:} $z=2$.}
\label{fig:oldKmin}
\end{figure}

In Fig.~\ref{fig:oldKmin} we present the results for the X-ray temperature function from the statistics of gravitational potential perturbations using the alternative definition of $k_\mathrm{min}$ and compare it to the Press-Schechter approach for three different cosmologies and two redshifts for temperatures between 1 and 30~keV. We can see clearly that both functions match quite well for very high temperatures, especially in the EdS case. For all three models, the temperature function derived from the gravitational potential is much flatter for low temperatures than the temperature function inferred from the Press-Schechter mass function so that the number density of objects with temperatures of about 1 keV is too low by a factor of more than 100.

This discrepancy can be explained considering that we expect many more density maxima than potential minima for the same volume of space due to the following reason. The Gaussian random field of potential perturbations is much smoother and has much more power on large scales than the corresponding field for the density contrast due to the the steepness of the potential power spectrum. Only large structures that have a high density contrast also correspond to a potential minimum, smaller structures only correspond to a maximum in the potential's Laplacian but not to a minimum in the potential itself. Since they are not located at a minimum of the potential, they have a non-vanishing potential gradient which corresponds to a non-zero peculiar velocity.

The definition presented in Eq.~(\ref{eq:kMin}) does not involve these problems due to the fact that large-scale potential gradients are removed and therefore, the condition $\vec{\eta}=\vec{0}$ is also applicable for structures with a low X-ray temperature.

\end{appendix}

\end{document}